%% file: main.tex
\documentclass[amsmath,amssymb,aps,prl,twocolumn,showpacs,floatfix,superscriptaddress,nofootinbib]{revtex4-2}
\usepackage[english]{babel}

\usepackage{graphicx}%
\usepackage{dcolumn}%
\usepackage{bm}%
\usepackage{physics}
\usepackage{hyperref}%
\usepackage{color}
\usepackage{ulem} %
\usepackage{multirow}
\usepackage{booktabs}
\usepackage{blkarray}
\usepackage{tabularray}
\usepackage[ruled]{algorithm2e} 
\usepackage[bb=dsserif]{mathalpha}

\newcommand{\I}{\mathcal{I}}
\newcommand{\F}{\mathcal{F}}

\begin{document}
\title{Belief Propagation-based Disentanglers for Tensor Network State Preparation}

\author{Tomasz Szo\l{}dra}
\affiliation{Zentrum für Optische Quantentechnologien, Fachbereich Physik, Universität Hamburg, Luruper Chaussee 149, 22761 Hamburg, Germany}
\author{Peter Schmelcher}
\affiliation{Zentrum für Optische Quantentechnologien, Fachbereich Physik, Universität Hamburg, Luruper Chaussee 149, 22761 Hamburg, Germany}

\date{\today}

\begin{abstract}
We develop a quantum circuit synthesis method for preparing a class of tensor network states. The scheme applies to states tractable with belief propagation (BP), a tensor network gauging scheme which recently allowed for classical simulations at large scales. The problem is reduced to independent, strictly local, classical variational optimizations: each nearest-neighbor two-qubit "disentangler" gate minimizes the entropy defined on an edge. Disentanglers drive the state to a product state and their Hermitian conjugate prepares the target. Each disentangling layer has depth at most $z+1$ (with $z$ the maximal number of nearest neighbors per site), the optimization has no barren plateaus, and the bond dimension stays bounded. As a demonstration, with only $3$-$5$ disentangling layers we prepare a $102$-qubit tree tensor network encoding a $17$-dimensional normal distribution and the transverse-field Ising model ground states on a $64$- to $127$-qubit heavy-hex lattice with fidelities of order $0.9-0.999$. The method opens new possibilities for quantum applications by transferring classical tensor network states onto hardware.
\end{abstract}

\maketitle

\begin{figure*}
	\centering
	\includegraphics{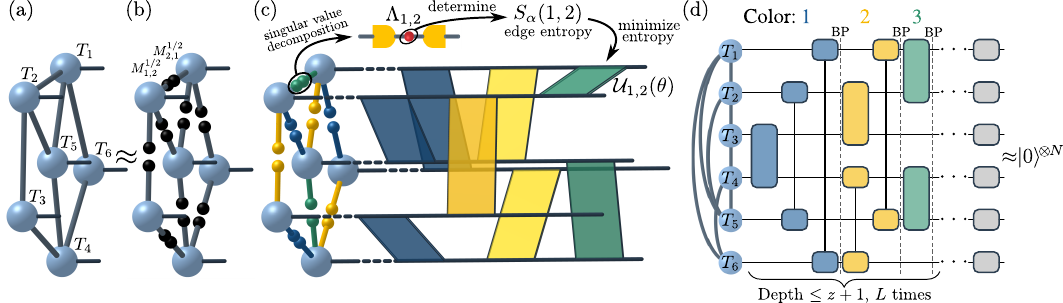}
	\caption{Belief Propagation Disentangler. (a) Tensor network state with one tensor per qubit. (b) Belief propagation approximation of (a): each tensor's environment is replaced with BP messages on incident edges (see text). (c) Edge entropy, Eq.~\eqref{eq:renyi}, is computed from the singular values of $M^{1/2}_{\mu,\nu}M^{1/2}_{\nu,\mu}$. Two-qubit disentangler gate applied around a bond minimizes the edge entropy and solves a strictly local variational problem. Bonds are colored with a minimal number of colors which do not repeat in a given node (here, 3 colors). Disentanglers are optimized and applied on the bonds of the same color in parallel. (d) Corresponding quantum circuit. Gates of the same color commute; BP is reiterated after each color.  After $L$ layers, single-qubit unitaries rotate the state towards $\ket{0}^{\otimes N}$. Hermitian conjugate of the circuit prepares the target.}
	\label{fig:idea}
\end{figure*}

Quantum state preparation (\textbf{QSP}) precedes execution of many quantum algorithms, for example Hamiltonian time evolution \cite{lloyd_universal_1996, miessen_quantum_2023}, ground state search \cite{peruzzo_variational_2014, cerezo_variational_2021}, computational quantum chemistry \cite{aspuruguzik_simulated_2005, mcardle_quantum_2020}, linear system solvers \cite{harrow_quantum_2009,yalovetzky_solving_2024}, and quantum machine learning \cite{schuld_introduction_2014,peralgarcia_systematic_2024}. While an exact preparation of arbitrary $N$-qubit states requires quantum resources exponential in $N$ \cite{shende_synthesis_2006,plesch_quantum_2011}, exploiting the structures present in the quantum state and allowing for an approximation error typically leads to more efficient QSP, essential for noisy hardware.

An example of such a structure is a tensor network (\textbf{TN}) form, a decomposition of high-dimensional tensors in terms of products of smaller tensors, which is a powerful ansatz for studies of quantum many-body systems \cite{banuls_tensor_2023, cirac_matrix_2021}. In one dimension, Matrix Product States (\textbf{MPS}) \cite{fannes_finitely_1992, schollwoeck_densitymatrix_2011} allow for polynomial-cost contractions, canonical form (local orthogonal bases), and well-conditioned algorithms, such as ground state search \cite{white_density_1992} and time evolution \cite{vidal_efficient_2003, haegeman_timedependent_2011,paeckel_timevolution_2019}. Several practical QSP methods have been developed for MPS \cite{schoen_sequential_2005,ran_encoding_2020,araujo_lowrank_2024,bendov_approximate_2024,rudolph_decomposition_2023,melnikov_quantum_2023,smith_constant_2024,malz_preparation_2024,mansuroglu_preparation_2026,green_quantum_2026,wei_state_2026,szoldra_scalable_2026,mingare_practical_2026, murota_exact_2026}, including hardware realizations \cite{smith_crossing_2022,zhang_qubit_2022,wall_quantum_2024,anselme_combining_2024,jaderberg_variational_2025,bohun_scalable_2026,scheer_renormalization_2025}. Generalization to Tree Tensor Networks (\textbf{TTN}) \cite{footnote1} inherits most of the favorable properties of MPS and remains polynomially contractible \cite{shi_classical_2006, murg_simulating_2010}. Some recent works have addressed the preparation of TTN states on near-term quantum hardware \cite{sugawara_embedding_2025, manabe_state_2025, ballarin_efficient_2025}.  

More broadly, Projected Entangled Pair States \cite{verstraete_renormalization_2004, verstraete_matrix_2008, cirac_matrix_2021} in two dimensions, and TN states on arbitrary graphs contain loops: they are hard to contract exactly \cite{schuch_computational_2007,markov_simulating_2008} and admit no exact canonical form. 
Existing preparation schemes for loopy networks rely on additional assumptions: a restricted class of tensors \cite{zaletel_isometric_2020, boesl_quantum_2025, schwarz_preparing_2012, schwarz_preparing_2013, wei_sequential_2022}, a local parent Hamiltonian with a unique ground state \cite{wei_efficient_2023}, or non-unitary resources such as measurement and feedforward \cite{sahay_finitedepth_2024, zhang_characterizing_2024} or dissipative engineering \cite{baruah_dissipative_2026}.
The variational route: globally optimizing the circuit fidelity with the target, as done for MPS and TTN \cite{melnikov_quantum_2023, shirakawa_automatic_2024, ballarin_efficient_2025}, requires starting from a good initialization, which is not always available, and evaluating that fidelity, which on a loopy graph cannot be done exactly at scale.
To our knowledge, no method prepares TN states on arbitrary bounded-degree graphs through a purely classical, barren-plateau-free optimization of a (practically shallow) digital quantum circuit.

Aimed at filling this gap, we introduce a QSP scheme applicable to TN states admitting a Belief Propagation (\textbf{BP}) gauge \cite{alkabetz_tensor_2021, tindall_gauging_2023}. The BP substitutes the rest of the network as seen from a given tensor by mean-field-like BP \emph{messages}. Being exact for MPS/TTN and approximate in the presence of loops, BP-based simulations recently overturned a claim of quantum utility on the IBM Eagle experiment \cite{tindall_efficient_2024, begusic_fast_2024}, and generalizations and systematic corrections to BP define the current state-of-the-art of TN simulations \cite{evenbly_loop_2026, gray_tensor_2026, midha_beyond_2026, midha_belief_2026, tindall_contracting_2026, alam_onset_2026}.
Our core idea is to disentangle the state of interest to a product state by applying variationally optimized nearest-neighbor gates around each bond in the TN, minimizing the local \emph{edge entropy} \cite{evenbly_gauge_2018,tindall_efficient_2024,watanabe_tensor_2026} computed from the BP messages; we refer to the method as Belief Propagation-based Disentangler (\textbf{BPD}). An optimal gate ordering ensures that multiple bonds can be independently disentangled in parallel, at a depth bounded by the TN coordination number, see Fig.~\ref{fig:idea}. The Hermitian conjugate of the total disentangling unitary prepares the target. BPD thus generalizes the Classical Variational Disentangler (\textbf{CVD}) algorithm for MPS \cite{mansuroglu_preparation_2026} to TTNs and loopy TNs.

\textit{Preliminaries.}---
\label{sec:preliminary}
We consider a TN state $\ket{\psi}$ with exactly one tensor $T_\mu$ per qubit but otherwise an arbitrary network structure, see~Fig.~\ref{fig:idea}(a), virtual index/bond dimension at most $\chi$, and $N$ qubits in total. The maximum number of a tensor's nearest neighbors (coordination number) is denoted by $z$.

Calculations with the state involve evaluating the norm $\braket{\psi}{\psi}$, which is an exponentially costly tensor contraction in general \cite{schuch_computational_2007,markov_simulating_2008} due to the presence of loops. Belief propagation allows for polynomial in $N$ contraction cost by replacing each $T_\nu$ tensor's environment (edges incident from other tensors $T_\mu$) by $\chi \times \chi$ \emph{message} matrices $M_{\mu,\nu}$, which satisfy the fixed-point BP equation
\begin{equation}
	M_{\nu, \mu'} = \left(\prod_{\mu\neq \mu'}M_{\mu,\nu}\right)\mathcal{T}_\nu,
	\label{eq:bp}
\end{equation}
 where $\mathcal{T}_\nu=\sum_s T^{(s)}_\nu{T}^{*(s)}_\nu$ and $s$ is the physical index. Equation \eqref{eq:bp} is a self-consistent relation between messages $M_{\mu,\nu}$ directed into $\mathcal{T}_\nu$ and messages $M_{\nu,\mu'}$ directed out from $\mathcal{T}_\nu$. It is solved iteratively, and is exact on trees, where it converges in a single iteration, and approximate for loopy networks.
In order to work with only the ket state $\ket{\psi}$, in contrast to the bra-ket object of Eq.~\eqref{eq:bp}, one puts on its bonds the message square roots $M^{1/2}$ \cite{footnote4}, see Fig.~\ref{fig:idea}(b), defined by $M^{1/2} = D^{1/2} U$, where $M=U^\dagger D U$ is the eigendecomposition and $M=(M^{1/2})^\dagger M^{1/2}$ \cite{tindall_gauging_2023}. 

BP not only allows one to approximately contract the norm and compute expectation values but also provides information about the entanglement properties of $\ket{\psi}$. On every bond the messages define an \emph{edge entropy}, which reduces to the ordinary bipartite entanglement entropy on a tree, generalizes it to loopy graphs, and vanishes for a product state, which we exploit in the disentangling scheme below. This is due to the fact that at the fixed point the site tensors with absorbed messages are approximately isometric and the BP gauge coincides with Vidal's gauge \cite{alkabetz_tensor_2021, tindall_gauging_2023}. In Vidal's gauge, bond singular values are by construction the Schmidt coefficients of the cut through edge $(\mu,\nu)$ on a tree. The same construction defines singular values on every bond of a general graph, yet without the same bipartite interpretation. In the BP framework, the singular values $\Lambda_{\mu,\nu}$ follow from a singular value decomposition (\textbf{SVD}) of the $\chi \times \chi$ product of messages \cite{tindall_gauging_2023}, Fig.~\ref{fig:idea}(c),
\begin{equation}
	M^{1/2}_{\mu,\nu}M^{1/2}_{\nu,\mu}=U_{\mu,\nu}\Lambda_{\mu,\nu}V^\dagger_{\mu,\nu}.
	\label{eq:svd}
\end{equation}
Normalizing $\Lambda_{\mu,\nu}$ so that $\sum_{k=1}^{\chi}(\Lambda_{\mu,\nu})^2_k=1$, the $\alpha$-R\'enyi entropy on an edge \cite{watanabe_tensor_2026} (here we generalize from $\alpha = 1$ to $\alpha>0$; see also \cite{evenbly_gauge_2018,tindall_efficient_2024}) reads 
\begin{equation}
	S_\alpha(\mu,\nu)=\frac{1}{1-\alpha}\log\sum_{k=1}^{\chi}(\Lambda_{\mu,\nu})^{2\alpha}_k,
	\label{eq:renyi}
\end{equation}
with the $\alpha\to1$ limit giving the von Neumann case. On a tree, Eq.~\eqref{eq:renyi} is directly the bipartite entanglement entropy for the cut going through the edge $(\mu,\nu)$; for loopy networks and $\alpha=1$ it is an approximation when summed over edges on a cut \cite{watanabe_tensor_2026}. Note that the edge entropy includes not just the two-body correlations between sites $(\mu,\nu)$ but also correlations at a finite range of other sites carried by the bond $(\mu,\nu)$.

\begin{figure}
	\includegraphics{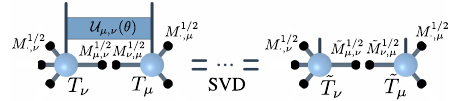}
	\caption{Gate application in the BP gauge updates only $T_{\mu}, T_{\nu}, M_{\mu,\nu}, M_{\nu,\mu}$ - see \cite{tindall_gauging_2023} for the complete algorithm.}
\label{fig:gate}
\end{figure}

In order to compute the action of a two-qubit gate on the physical indices of $(\mu, \nu)$ on the state, one uses only the tensors $T_\mu$, $T_\nu$ and their associated BP messages \cite{tindall_gauging_2023}. Gate application modifies $T_\mu$, $T_\nu$ and the two messages $M_{\mu,\nu}$, $M_{\nu,\mu}$, leaving all other messages invariant, see Fig.~\ref{fig:gate}. Gate's influence on the $(\mu,\nu)$ edge entropy is thus captured by a strictly local tensor network object.

\textit{BPD state preparation}.--- 
Our aim is to drive the target state $\ket{\psi}$ to a product state, hence on every bond we look for the \emph{disentangler} gate minimizing the edge entropy after its application. The disentangler is parametrized~as~\cite{mansuroglu_preparation_2026}
\begin{align}
	\mathcal{U}_{\mu,\nu}(\theta)={}&e^{-i\theta_1 X_\mu X_\nu}e^{-i\theta_2 Y_\mu Y_\nu}e^{-i\theta_3 Z_\mu Z_\nu}\nonumber\\
	&\times\left(e^{-i(\theta_4 X_\nu+\theta_5 Y_\nu+\theta_6 Z_\nu)}\right.\nonumber\\
	&\qquad\otimes\left.e^{-i(\theta_7 X_\mu+\theta_8 Y_\mu+\theta_9 Z_\mu)}\right),
	\label{eq:general_disentangler}
\end{align}
which is a decomposition of a general two-qubit SU(4) matrix \cite{kraus_optimal_2001} without the pair of single-qubit gates at the end, which do not influence the entanglement; hardware-native gates could be used if desired. 
We denote $S_\alpha(\mu,\nu; \theta)$ as the edge entropy after applying gate $\mathcal{U}_{\mu,\nu}(\theta)$. Minimizing $S_\alpha(\mu,\nu; \theta)$ over the nine angles $\theta$ is a strictly local variational problem: the cost function and its gradient involve only $T_\mu$, $T_\nu$ and their incident messages. There are no gradients vanishing exponentially with $N$, also known as barren plateaus \cite{mcclean_barren_2018, cerezo_cost_2021}, that typically appear in global variational optimizations of quantum circuits. Generalizing the 2-R\'enyi entropy MPS calculation from Ref.~\cite{mansuroglu_preparation_2026}, in Supplemental Material (SM) \cite{supplemental} we show that assuming a (Haar-)random neighborhood of the two sites $\mu,\nu$, at $\theta=0$, the expected value of the gradient along direction $i$, corresponding to any pair of Pauli operators from $\sigma_\nu \otimes \sigma_\mu \in \lbrace X\otimes X, Y\otimes Y, Z \otimes Z\rbrace$, vanishes, $\mathbb{E} \left[ \left. \partial_i S_2(\mu,\nu; \theta)\right|_{\theta=0}\right] = 0$. The variance reads 
\begin{equation}
	\mathbb{E} \left[\left. \left(\partial_i S_2(\mu,\nu;\theta)\right)^2\right|_{\theta=0}\right] = \frac{8 M_\mu M_\nu \left( e^{2(S_2 - S_3)} -1\right)}{(M_\mu^2-1)(M_\nu^2-1)} ,
	\label{eq:variance}
\end{equation}
where $M_\mu=2\chi^{z_\mu-1}$, $M_\nu=2\chi^{z_\nu-1}$, and $z_\mu,z_\nu\leq z$ denote the number of nearest neighbors of sites $\mu,\nu$. Thus, the gradient variance decreases only polynomially with the bond dimension $\chi$, and the system size $N$ does not enter Eq.~\eqref{eq:variance}. We use autodifferentiation \cite{footnote2} for gradients and employ the L-BFGS-B \cite{zhu_lbfgsb_1997} algorithm for optimization. 

Two disentanglers can be independently optimized and applied simultaneously if they do not share a node. Scheduling all edges into parallel rounds is the \emph{edge coloring problem}, routinely met also in e.g. trotterized time evolution gate ordering \cite{rudolph_simulating_2025}. A coloring with $K\leq z+1$ colors always exists and is efficiently constructible \cite{misra_constructive_1992}, giving a single disentangling layer of depth $K\leq z+1$. For some graphs, such as bipartite graphs \cite{footnote3}, $K=z$ colors suffice, see End Matter. Each color makes one round of commuting nearest-neighbor gates, Fig.~\ref{fig:idea}(c), after which BP can be reiterated to restore the gauge. The special case of a brick-wall CVD circuit for MPS \cite{mansuroglu_preparation_2026} with $z=2$ is recovered with $K=2$ colors for even/odd bonds.

After $L$ layers, single-qubit unitaries rotate each site towards $\ket{0}$, see Fig.~\ref{fig:idea}(d). The gates follow from truncating the BP-gauge bond dimensions \cite{tindall_gauging_2023,gray_quimb_2018} to $\chi=1$, interpreting the product state as an MPS, and applying a standard MPS-to-circuit mapping \cite{schoen_sequential_2005}. The Hermitian conjugate of the circuit prepares the target from $\ket{0}^{\otimes N}$.

Classical efficiency guarantee established for CVD \cite{mansuroglu_preparation_2026} carries over. If the singular values discarded at the subsequent truncation carry total weight ${p=\sum_{k>\chi_{\mu,\nu}}(\Lambda_{\mu,\nu})^2_k}$, the retained rank obeys
\begin{equation}
	\chi_{\mu,\nu}\leq (1-\alpha)\,\alpha^{\frac{\alpha}{1-\alpha}}\,p^{-\frac{\alpha}{1-\alpha}}\,e^{S_\alpha(\mu,\nu)},
	\label{eq:bond_bound}
\end{equation}
for any $\alpha\in(0,1)$, which is Lemma 1 of Ref.~\cite{mansuroglu_preparation_2026} read on an edge, see End Matter. As in \cite{szoldra_scalable_2026}, throughout this work we use $\alpha=2$ (without the efficiency guarantee, but with a better numerical stability) and we still find the bond dimension to stay bounded throughout the disentangling.

\begin{figure}
	\includegraphics[width=.9\columnwidth]{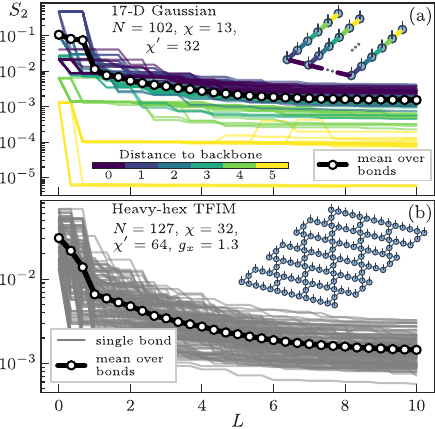}
	\caption{Edge 2-R\'enyi entropy as a function of the number of disentangling layers for (a) multivariate normal distribution and (b) ground state of the TFIM defined on the heavy-hex topology. Application of disentangling unitaries systematically decreases the bond entropy.}
	\label{fig:disentangling}
\end{figure}

\textit{Results}.---
\label{sec:results}
Our first target is the $n$-dimensional normal distribution
\begin{equation}
	f(\bm{x};\bm{\mu},\Sigma)\propto\exp\left(-\frac{1}{2}(\bm{x}-\bm{\mu})^{\rm T}\Sigma^{-1}(\bm{x}-\bm{\mu})\right),
	\label{eq:gaussian}
\end{equation}
on $\bm{x}\in[0,1)^n$, amplitude-encoded as $\ket{\psi}\propto\sum_{\bm{x}}f(\bm{x})\ket{\bm{x}}$, whose loading is a recurring task \cite{rosenkranz_quantum_2025,manabe_state_2025, ballarin_efficient_2025}.  
Following \cite{ballarin_efficient_2025}, we take $n=17$ variables, $\mu_i=0.5$, and $\Sigma$ tridiagonal with $\Sigma_{ii}=0.05$ and $\Sigma_{i,i\pm1}=0.01$. Each variable is discretized on $2^{n_x}$ points in the quantics \cite{oseledets_approximation_2009, khoromskij_quantics_2011, ritter_quantics_2024} representation $x_i=\sum_{k=1}^{n_x}s_{i,k}2^{-k}$, $s_{i,k}\in\{0,1\}$, with $n_x=6$, so that the $k$-th qubit of $x_i$ resolves the lengthscale $2^{-k}$, giving $N=n\,n_x=102$ qubits. The TTN is built by tensor cross interpolation \cite{ballani_black_2013, tindall_compressing_2024, fernandez_learning_2025}, on the comb geometry of Fig.~\ref{fig:disentangling}(a): each variable's $n_x$ qubits chain from the coarsest scale at the backbone to the finest at the leaf; see End Matter for details. This gives a TTN with $\chi=13$, requiring only $\approx5.5\cdot 10^5$ evaluations of $f(\bm{x})$.

The second target is the ground state of the transverse-field Ising model (\textbf{TFIM}) 
\begin{equation}
	H=-\sum_{\langle i,j\rangle}Z_iZ_j-g_x\sum_{j}X_j,
	\label{eq:H_tfim}
\end{equation}
on the heavy-hexagonal lattice of the $127$-qubit IBM Eagle processor \cite{kim_evidence_2023}, Fig.~\ref{fig:disentangling}(b), whose critical point lies around $g_x\approx1.4-1.5$ \cite{lin_utilityscale_2026, martin_preoptimization_2026}. We obtain the state by imaginary-time evolution in the BP gauge; see SM for details and a comparison to alternative methods \cite{lin_utilityscale_2026}.

\begin{figure*}
	\includegraphics{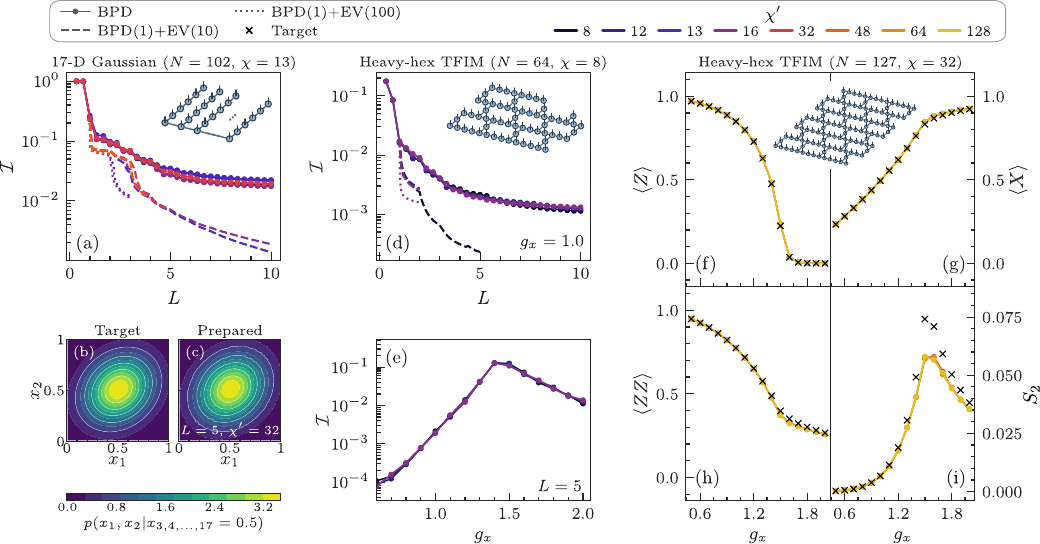}
	\caption{(a) Infidelity of the 17-D Gaussian Eq.~\eqref{eq:gaussian} saturating close to $\mathcal{I}\approx 0.02$ after $L=5$ BPD layers. With EV sweeps one can reach better infidelities. (b-c) Probability distribution of variables $x_1, x_2$ with other variables fixed to $\mu=0.5$ for (b) target and (c) circuit from BPD(1)+EV(10). Positive correlation from (b) is almost ideally reproduced in (c). (d) Same as (a) but for the ground state of the TFIM on heavy-hex lattice restricted to 64 qubits, Eq.~\eqref{eq:H_tfim}, $g_x=1.0$, with a small $\chi=8$, $\chi'\leq 16$ so that the overlap network is exactly contractible. (e) Infidelity across the phase diagram, peaking around $g_x\approx 1.4-1.5$. (f-i) Properties of the 127-qubit TFIM ground state. Deviations from the target again occur close to $g_x\approx 1.4-1.5$, where BP is not accurate. Panel (i) shows that the average edge entropy at bonds peaks around the critical point and is not fully reproduced by the BPD circuit. Datapoints for $\chi'=32,48,64,128$ mostly overlap, with the $\chi'=128$ visible on top.
	}
	\label{fig:main}
\end{figure*}
Both geometries are bipartite, so edge coloring yields $K=z=3$ colors. Figure~\ref{fig:disentangling} shows the edge 2-R\'enyi entropy of every bond vs. the number of BPD layers $L$.  Here, $\chi'$ denotes BPD simulation's bond dimension, in contrast to the target's $\chi$. For the Gaussian, Fig.~\ref{fig:disentangling}(a), the initial entropy values form bands set by distance from the comb backbone. Bonds at large distances, linking to qubits specifying $x_i$ at the smallest lengthscales $2^{-5}, 2^{-6}$, which are below the Gaussian width, carry an exponentially small entanglement due to the function smoothness \cite{bohun_scalable_2026}. One BPD layer already removes most of the entropy, and further layers reduce the mean until saturation, a consequence of a bond-by-bond heuristic. In Fig.~\ref{fig:disentangling}(b), for the TFIM in the "hard" regime $g_x=1.3$ near criticality (long-ranged correlations), individual and mean edge entropies keep decreasing up to $L=10$. In both cases, we are left with the edge entropies very close to zero. We also investigated the disentangling influence on the BP error, i.e. whether the loop contributions increase, and found that the BP error remains at a stable level close to initial value, see End Matter for details and definitions.

A decreasing edge entropy does not by itself certify a good preparation, so we turn to the infidelity $\I=1-\F$, where $\F=|\braket{\psi_{\rm prep}}{\psi}|^2$ is the fidelity of the \emph{prepared} state (after truncations to $\chi'$) $\ket{\psi_{\rm prep}}$. This is in contrast to fidelity of the disentangled state with $\ket{0}^{\otimes N}$; the two coincide for sufficient $\chi'$, which was verified for the presented cases. For the Gaussian, Fig.~\ref{fig:main}(a), a single BPD layer brings the infidelity from $\I\approx1$ down to $\I\approx0.2$, and further layers reduce it to $\I\approx0.02$, where the heuristic saturates. We further refine the circuit with the Evenbly-Vidal (\textbf{EV}) sweeping gate optimizer \cite{evenbly_algorithms_2009, shirakawa_automatic_2024,rudolph_decomposition_2023,szoldra_scalable_2026}, requiring a computation of the "environment tensor" for each gate, roughly of the same complexity as contracting the overlap network, hence tractable for a TTN, see End Matter. It is applied after every BPD layer on the whole circuit in $n$ sweeps, denoted by BPD(1)+EV($n$). We observe that BPD(1)+EV(10) reaches $\I<0.01$ at $L=5$, while the more expensive BPD(1)+EV(100) attains a comparable value already at $L=3$.
In comparison, Ref.~\cite{ballarin_efficient_2025} reaches an infidelity $\I = 0.0043$ at the same circuit depth for $L=3$, but, unlike our scheme, it requires an interpolation from an easy-to-prepare state to the target, and a manually designed circuit ansatz. Figure~\ref{fig:main}(b-c) shows the preparation quality more directly: the distribution of $x_1,x_2$ with the remaining fifteen variables fixed at $x_i=\mu=0.5$. The tridiagonal $\Sigma$ induces a small positive correlation between $x_1$ and $x_2$, visible as tilted ellipse contours in the target, Fig.~\ref{fig:main}(b), reproduced perfectly by the BPD(1)+EV(10) method at $L=5$, $\I\approx0.01$, Fig.~\ref{fig:main}(c).

For the TFIM we first choose a controllable setting where the infidelity can be computed by exact contraction \cite{gray_hyper_2021}: $N=64$ qubits and $\chi=8$, $\chi'\leq16$. Gates are still applied in the BP gauge. At $g_x=1.0$, two layers suffice for $\I<0.01$, and the heuristic saturates around $\I\approx10^{-3}$ beyond $L\gtrapprox5$, Fig.~\ref{fig:main}(d); combination with EV sweeps gives $\I<10^{-3}$ with only $2$-$3$ layers. Computing infidelity for larger $L$ would require an approximate contraction scheme, e.g. a boundary MPS method \cite{tindall_efficient_2024, rudolph_simulating_2025}. Scanning the phase diagram in Fig.~\ref{fig:main}(e), the infidelity for 5 layers makes a smooth function of $g_x$, drops to $\I\approx10^{-4}$ deep in the ferromagnetic phase, and peaks at $\I\approx0.1$ close to the critical point $g_x\approx1.4-1.5$. This is expected: correlations are long-ranged at criticality, whereas our circuit has a finite lightcone \cite{lieb_finite_1972,bravo_prieto_scaling_2020}. Away from the transition, preparation is accurate, and the smooth $g_x$ dependence indicates stable behavior of the method.

Finally, we benchmark BPD where the exact contraction is out of reach: $N=127$ qubits and $\chi=32$, $L=5$ BPD layers, Fig.~\ref{fig:main}(f-i). Observables are evaluated within the BP approximation \cite{tindall_gauging_2023}, both in the prepared and target states. The order parameter $\langle Z\rangle$, transverse magnetization $\langle X\rangle$ and nearest-neighbor correlator $\langle ZZ\rangle$ of the prepared state, averaged over the lattice, follow the target across the whole phase diagram, with the error growing close to the transition, as expected. The curves for $\chi'=32,\dots,128$ are nearly indistinguishable, so that $\chi'=\mathcal{O}(\chi)$ suffices here. The edge entropy is most affected by criticality, Fig.~\ref{fig:main}(i): the error peaks at the transition, consistent with the infidelity peak in Fig.~\ref{fig:main}(e).

\textit{Conclusions and outlook}.---
\label{sec:conclusions}
We have introduced Belief Propagation-based Disentangler, a heuristic scheme preparing TN states on arbitrary graphs, including those with loops, as shallow circuits of nearest-neighbor gates. Every gate follows from a strictly local, entirely classical and barren-plateau-free minimization of the edge entropy, edge coloring arranges one layer of gates into parallel rounds, and the scheme can fully utilize the native hardware connectivity. Thus, we generalize the brick-wall MPS disentanglers \cite{mansuroglu_preparation_2026} to trees and loopy TNs. The method can be combined with fidelity optimizers such as the Evenbly-Vidal algorithm \cite{evenbly_algorithms_2009, shirakawa_automatic_2024,rudolph_decomposition_2023,szoldra_scalable_2026}. It also does not require a smooth interpolation to an easy-to-prepare target \cite{ballarin_efficient_2025}, or the target being the ground state of a known physical Hamiltonian \cite{lin_utilityscale_2026, martin_preoptimization_2026}.

Certain limitations set the scope of the method. As for any heuristic, there is no guarantee of obtaining a (shallow) accurate circuit for all targets. The number of layers determines each qubit's lightcone, and sites not connected by the lightcone will not be disentangled. Our results hold under the assumption that the BP gauge is a good approximation. Loosely speaking, the target state needs to have tree-like correlations and only long loops, such as in the heavy-hex topology. Rigorous conditions for BP applicability including systematic corrections are established in Ref.~\cite{midha_belief_2026}. 

Direct applications of the scheme include warm-starting hardware ground state search \cite{rudolph_synergistic_2023, martin_preoptimization_2026} or simulating circuits until classical tractability is lost and handing it over to the device, extending the total accessible circuit depth \cite{robertson_approximate_2025,jaderberg_variational_2025}. Moreover, the tool allows for loading classical data whose efficient TN description is beyond the MPS form \cite{tindall_compressing_2024, ballarin_efficient_2025}.

Finally, we note that higher-order corrections to BP \cite{evenbly_loop_2026, gray_tensor_2026, midha_beyond_2026, midha_belief_2026, tindall_contracting_2026, alam_onset_2026} widen the regime where the gauge is accurate and could enter the disentangling scheme. A related question concerns the EV sweeping optimizer, which we run only where the overlap network is exactly contractible; running it within the BP approximation would complete the toolbox.

\textit{Acknowledgements.}---
We acknowledge the funding of the Federal Ministry of Research, Technology and Space Project Number 13N17157 (Q-ROM). We thank Z. Zeybek for providing his useful suggestions and valuable comments on the manuscript.
We also thank C.-J. Lin for providing the numerical data from Ref.~\cite{lin_utilityscale_2026}. We acknowledge the use of open source software: \verb|quimb| \cite{gray_quimb_2018}, \verb|cotengra| \cite{gray_hyper_2021}, \verb|xfac| \cite{fernandez_learning_2025, xfac}.
The computational resources were provided by the PHYSnet-Rechenzentrum of Universität Hamburg.

\input main.bbl
\clearpage

\onecolumngrid
\begin{center}\large \textbf{End Matter}\end{center}
\twocolumngrid

\textit{Bounded bond dimension}.---  
Let us justify the bound on the bond dimension in Eq.~\eqref{eq:bond_bound}. Since $M_{\mu,\nu}$ is positive semidefinite at the fixed point \cite{tindall_gauging_2023}, its square root in Eq.~\eqref{eq:svd} exists and $(\Lambda_{\mu,\nu})_k^2$ is real and non-negative; the normalization $\sum_k(\Lambda_{\mu,\nu})^2_k=1$ makes it a probability distribution.
This is the only property of the squared Schmidt coefficients used in the proof of Lemma 1 of Ref.~\cite{mansuroglu_preparation_2026} (a bound on $\sum_i p_i^\alpha$ depending only on $p_i\geq0$, $\sum_i p_i=1$, and the discarded weight $p$), so substituting $(\Lambda_{\mu,\nu})_k^2$ for $p_i$ and setting $p=\sum_{k>\chi_{\mu,\nu}}(\Lambda_{\mu,\nu})^2_k$ reproduces that proof and Eq.~\eqref{eq:bond_bound} exactly on every edge.

\textit{Graph edge coloring}.---
Graph edge coloring is a problem of assigning a minimal number of $K$ colors to all edges on the graph, such that no two incident edges have the same color, i.e. colors of edges of any node are all different. By Vizing's theorem \cite{vizing_estimate_1964}, for a graph of maximum degree $z$, the number of colors satisfies $K\in\{z,z+1\}$, with $K=z$ reached for bipartite graphs \cite{koenig_uber_1916,koenig_english_2024,footnote3}, in particular trees. Edge coloring with at most $K=z+1$ colors can always be constructed in time polynomial in $N$ and the number of edges \cite{misra_constructive_1992}, but deciding whether a lower $K=z$ coloring exists is an NP-complete problem on arbitrary graphs \cite{holyer_npcompleteness_1981}.

\textit{Tensor cross interpolation}.---
Tensor cross interpolation (\textbf{TCI}) \cite{ballani_black_2013, tindall_compressing_2024, fernandez_learning_2025} allows us to construct the tree tensor network encoding the $n=17$-dimensional normal distribution, Eq.~\eqref{eq:gaussian} in the main text, without evaluating the function $f(\bm{x})$ on all possible $2^N=2^{102}$ arguments. The TCI algorithm uses only a selected subset of arguments that are sufficient to approximate the finite-rank TTN to a prescribed accuracy.
 \begin{table}[h]
 	\centering
 	\begin{tabular}{ll}
 		\toprule
 		Parameter & Value \\
 		\midrule
 		Relative tolerance & $10^{-12}$ \\
 		Bond dimension $\chi$ & $13$ \\
 		Initial pivot & $\bm{x}=\bm{\mu}$ \\
 		"DMRG" Sweeps & $10$ \\
 		\bottomrule
 	\end{tabular}
 	\caption{Tensor cross interpolation hyperparameters used to build the multivariate Gaussian comb TTN.}
 	\label{tab:tci_hyperparams}
 \end{table}To perform TCI we use the \verb|xfac| \cite{fernandez_learning_2025,xfac} package with hyperparameters set as in Table~\ref{tab:tci_hyperparams}. The total number of function $f(\bm{x})$ queries is $547606$. By evaluating the TTN on (i) $1000$ randomly sampled points in $[0,1)^n$, (ii) points of with coordinates close to the mean chosen as a set $\lbrace\bm{x}^{(k)}\rbrace_{k=1...17}$ where $x^{(k)}_j = \mu_j + 0.1\delta_{jk} - 0.05 (1-\delta_{jk})$, we verify that the obtained amplitudes match the original $f(\bm{x})$ with a maximal $5 \cdot 10^{-13}$ absolute error. 

\textit{Evenbly-Vidal optimization.}---
Writing the circuit as a product of $M$ gates $U=U_M\dots U_1$ and the overlap as $F=\bra{0}^{\otimes N}U^\dagger\ket{\psi}$, EV updates one gate $U_m$ at a time with all others fixed. Its environment $\mathcal{F}_m$ is the overlap network with $U_m$ removed, leaving two bra and two ket open legs, i.e. a $4\times4$ matrix; the optimal update maximizing $|F|$ is $U_m'=XY$, where $\mathcal{F}_m=XDY$ is the singular value decomposition \cite{shirakawa_automatic_2024}. Visiting all gates forward and backward constitutes a sweep.
\begin{figure}
	\includegraphics{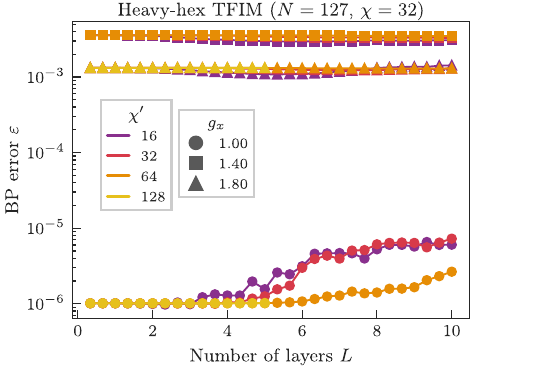}
	\caption{BP Error of the disentangled state as a function of the number of disentangling layers $L$. }
	\label{fig:bperror}
\end{figure}

\textit{BP error during disentangling}.---
In order to investigate whether edge disentangling doesn't degrade the subsequent belief propagation accuracy, after each layer of disentangling we measure the BP error $\varepsilon$ following the approximate estimation method of Ref.~\cite{rudolph_simulating_2025}. The error is computed from the spectrum of the transfer matrices corresponding to loops of the smallest size in the system (in our case of a heavy-hex lattice, loops of $12$ sites). For the norm network $\bra{\psi}\ket{\psi}$, for each such loop $\ell$, we absorb the BP messages incoming to the loop, cut one virtual bond, exactly contract the resulting object obtaining a transfer matrix of size $\chi^2 \times \chi^2$, and compute its eigenvalues as $\lambda^{\ell}_1, \lambda^{\ell}_2,...$, ordered by their decreasing absolute value. The error is defined as 
\begin{align}
	\varepsilon &= \frac{1}{N_{\text{loops}}}\sum_{\ell = 1}^{N_{\text{loops}}}\varepsilon_{\ell} \label{eq:avBPError}\\
	\varepsilon_{\ell} &= 1 - \frac{\vert \lambda^{\ell}_{1} \vert}{\sum_{i} \vert \lambda^{l}_{i} \vert} \label{eq:loopBPerror} ,
\end{align}
where the average is taken over all loops of the smallest size in the network. Loosely speaking, the error quantifies loop correlations which cannot be captured by BP, and where systematic loop corrections become necessary. The error satisfies $0 \leq \varepsilon \leq 1-1/\chi^{2}$. 

Figure \ref{fig:bperror} shows the BP error $\varepsilon$ of the disentangled TFIM state as a function of the number of disentangling layers (each point corresponds to a single color of gates, so there are three points per one layer). Clearly, the target state at $g_x=1.0$ has three orders of magnitude smaller initial BP error than at $g_x=1.4$, located closer to the critical point, which is consistent with the failure of BP at critical points \cite{midha_belief_2026}. Nevertheless, in all cases of $g_x\in\lbrace 1.0, 1.4, 1.8 \rbrace$, the error is stabilized close to the initial value, and does not uncontrollably grow when adding more disentangling layers. A small growth is observed for $g_x=1.0$, which is still low in the absolute value - increasing the bond dimension removes the issue. Overall, we find that applying local disentanglers to the state does not silently spoil the BP approximation.

\textit{BPD hyperparameters}.---
Table~\ref{tab:bpd_hyperparams} lists the hyperparameters common to both the Gaussian and TFIM disentangling runs of the main text. The "Pruning: singular value cutoff" sets the threshold for singular values such that when truncation at this level leaves a rank-1 decomposition, the bond is treated as disentangled, and further gate applications on that bond are discontinued to save on the total gate count.

\begin{table}
	\centering
	\begin{tabular}{ll}
		\toprule
		Parameter & Value \\
		\midrule
		BP Eq.~\eqref{eq:bp} relative tolerance & $10^{-12}$ \\
		BP Eq.~\eqref{eq:bp} max.\ iterations & $1000$ \\
		R\'enyi index $\alpha$ & 2\\
		Bond truncation cutoff & $10^{-14}$ \\
		Pruning: singular value cutoff & $10^{-4}$ \\
		L-BFGS-B tolerance & $10^{-10}$ \\
		L-BFGS-B max.\ iterations & $1000$ \\
		BP re-run interval (TTN) & None (tree graph) \\
		BP re-run interval (Heavy-hex) & After every gate color\\
		\bottomrule
	\end{tabular}
	\caption{Hyperparameters used for BP gauging and BPD gate optimization.}
	\label{tab:bpd_hyperparams}
\end{table}

\end{document}


\title{Supplemental Material for\\Belief Propagation-based Disentanglers for Tensor Network State Preparation}

\author{Tomasz Szo\l{}dra}
\affiliation{Zentrum für Optische Quantentechnologien, Fachbereich Physik, Universität Hamburg, Luruper Chaussee 149, 22761 Hamburg, Germany}
\author{Peter Schmelcher}
\affiliation{Zentrum für Optische Quantentechnologien, Fachbereich Physik, Universität Hamburg, Luruper Chaussee 149, 22761 Hamburg, Germany}

	\date{\today}

	\maketitle
	
	\setcounter{equation}{0}
	\setcounter{figure}{0}
	\setcounter{table}{0}
	\makeatletter
	\renewcommand{\theequation}{S\arabic{equation}}
	\renewcommand{\thefigure}{S\arabic{figure}}
	\renewcommand{\thetable}{S\arabic{table}}
	\renewcommand{\bibnumfmt}[1]{[S#1]}

	\section{Gradient of the edge entropy}
	Here we perform the calculation of $2$-R\'enyi entropy gradient $\left. \partial_i S_2(\theta)\right|_{\theta=0}$, its variance, and explicitly demonstrate the lack of barren plateaus during optimization. 
	To that end, we will use the more convenient TN representation in Vidal's gauge, which is equivalent to BP gauge, see \cite{tindall_gauging_2023} for details how to transform between these two gauges.
	
	We consider the tensor $\Gamma_\nu$ on site $\nu$ with all singular values $\Lambda_{\nu_i,\nu}$ (diagonal tensors) absorbed except $\Lambda_{\mu,\nu}\equiv \Lambda$, defining an isometry $A$:
	\begin{equation}
		\includegraphics{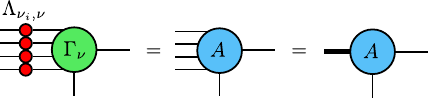}
		\label{eq:Adef}
	\end{equation}
	where for convenience with thick line we have defined a multi-index of dimension $\chi^{z_\nu-1}$, where $z_\nu \leq z$ is the number of nearest neighbors of site $\nu$. The isometry condition of Vidal's gauge gives
	\begin{equation}
		\includegraphics{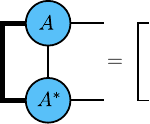}
	\end{equation}
	Similarly, for site $\mu$, 
	\begin{equation}
		\includegraphics{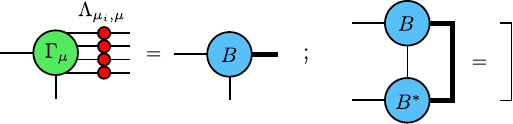}
	\end{equation}
	Definition of the edge entropy $S_2(\mu,\nu)$ in terms of tensor network expressions requires a special care in comparison to the usual bipartite case, e.g. for MPS, where the reduced density matrix, undefined for a cut bond on a loopy system, was originally used in \cite{mansuroglu_preparation_2026}. Thus, let us define $P=A \Lambda B$:
	\begin{equation}
		\includegraphics{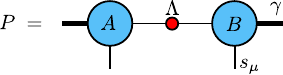}
	\end{equation}
	where we marked the $s_\mu$ physical index and $\gamma$ virtual index explicitly. Now, we define $Q = \Tr_{s_\mu, \gamma} (P P^\dagger)$:
	\begin{equation}
		\includegraphics{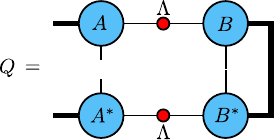}
	\end{equation}
	$Q$ is a substitue for the reduced density matrix of the "left subsystem" containing site $\nu$, but is well-defined in the presence of loops. Note that 
	\begin{equation}
		\includegraphics{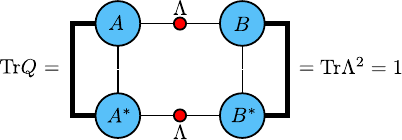}
	\end{equation}
	We now make an observation that $S_2(\mu,\nu) = -\log \Tr(\Lambda^4) = -\log \Tr Q^2$. This is due to 
	\begin{equation}
		\includegraphics{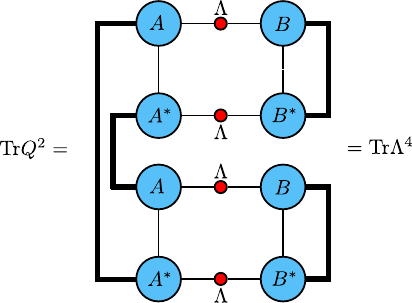}
	\end{equation}
	which follows from isometry properties of $A$, $B$. Let us now determine $S_2(\mu, \nu; \theta)$ after applying gate $\U_{\mu,\nu}(\theta)$ on sites $\mu,\nu$. For brevity, we write $S_2(\mu, \nu; \theta)\equiv S_2(\theta)$ and $\U_{\mu,\nu}(\theta)\equiv \U(\theta)$. Gate application modifies $P$ to $P(\theta) = \U(\theta) A \Lambda B$,
	\begin{equation}
		\includegraphics{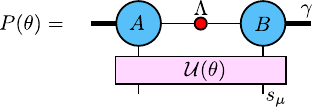}
	\end{equation}
	Similarly, $Q$ transforms to $Q(\theta) = \Tr_{s_\mu, \gamma} \left(P(\theta) P^\dagger(\theta)\right)$,
	\begin{equation}
		\includegraphics{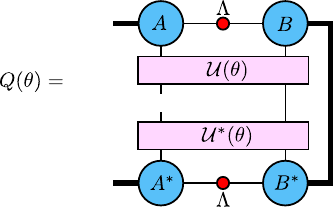}
	\end{equation}
	and $S_2(\theta) = -\log \Tr (Q(\theta)^2)$.
	
	Let us compute the gradient $\partial_i S_2(\theta) \equiv \partial S_2(\theta) / \partial_{\theta_i}$:
	\begin{equation}
		\partial_i S_2(\theta) = -\frac{2}{\Tr (Q(\theta)^2)} \Tr ( (\partial_i Q(\theta)) Q(\theta)).
	\end{equation}
	Note that 
	\begin{equation}
		\partial_i Q(\theta) =  \Tr_{s_\mu, \gamma} \left[ (\partial_i P(\theta)) P^\dagger(\theta) \right] + \Tr_{s_\mu, \gamma} \left[ P(\theta) (\partial_i P^\dagger (\theta))  \right].
	\end{equation}
	Considering $i$ for $\sigma_\nu \otimes \sigma_\mu \in \lbrace X\otimes X, Y\otimes Y, Z \otimes Z\rbrace$ (i.e. two-qubit generators for $\U(\theta)$), at $\theta=0$:
	\begin{equation}
		\left. \partial_i P(\theta) \right|_{\theta=0} = \left.\left(\partial_i \U(\theta) \right)\right|_{\theta=0} A \Lambda B = -i (\sigma_\nu \otimes \sigma_\mu) P,
	\end{equation}
or graphically
	\begin{equation}
		\includegraphics{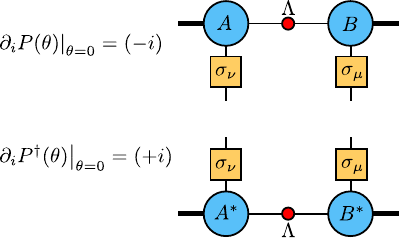}
	\end{equation}
	Thus, 
	\begin{equation}
		\includegraphics{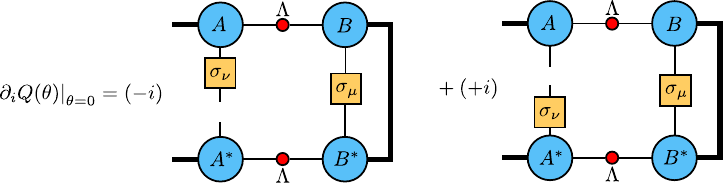}
	\end{equation}
	Finally, 
	\begin{equation}
		\includegraphics{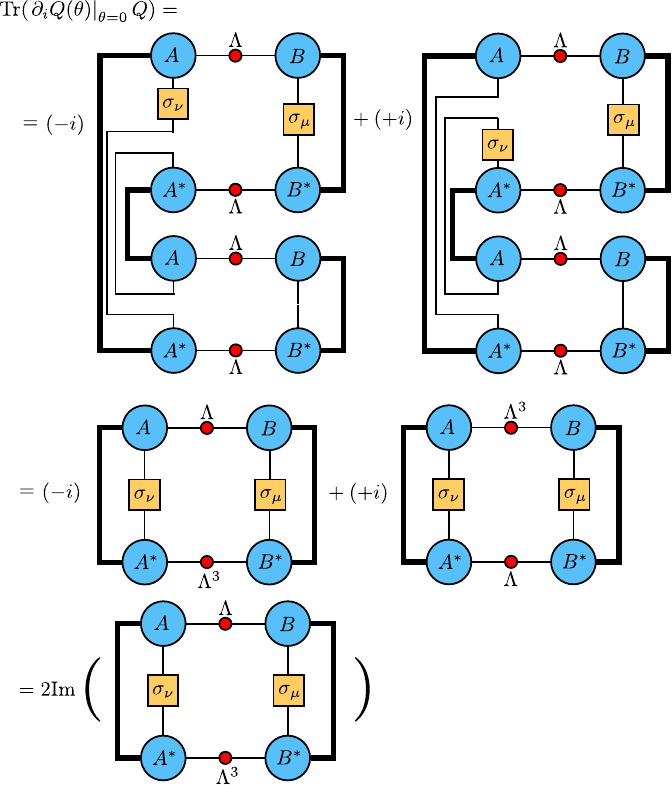}
	\end{equation}
	where we again used the isometry properties of $A$, $B$. The gradient has a form fully analogous to the MPS case \cite{mansuroglu_preparation_2026} except for different dimensionalities of the isometries $M_A=2\chi^{z_{\nu}-1}$ for $A$ and $M_B=2\chi^{z_{\nu}-1}$ for $B$, and not $M_A=M_B=2\chi$ as for the MPS. The gradient is identically equal to zero for a product state $\Lambda=1$, or when $\Lambda=\lambda \mathbb{1}$ (evenly distributed signular values make the expression under $\Im(\dots)$ proportional to $\lambda^4$ and invariant under complex conjugate), or in special cases when all choices of $\sigma_\nu \otimes \sigma_\mu$ yield real value in expression under $\Im(\dots)$.
	
	To investigate the behavior of the gradient for generic states, let us consider the statistical expectation value of the gradient and its variance under the assumption that isometries $A$, $B$ are columns of unitaries $U\in SU(M_A)$, $V \in SU(M_B)$:
	\begin{equation}
		\includegraphics{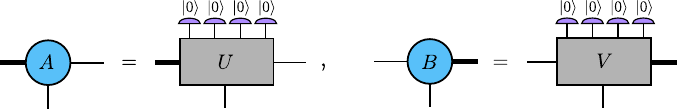}
	\end{equation}
	with $U, V$ Haar-random distributed and independent. We stress that at this point the independence of $A$ and $B$ is assumed, which holds only if we neglect the  influence of loops as a higher-order correction. This is approximately satisfied for the considered BP-gaugeable TN states. The calculation based on the Weingarten calculus then follows Ref.~\cite{mansuroglu_preparation_2026} verbatim upon replacement of the unitary dimensionalities. The only difference is the numerical prefactor, so we do not repeat the calculation here. The final result reads
	\begin{equation}
		\mathbb{E} \left[ \left. \partial_i S_2(\theta)\right|_{\theta=0}\right] = \frac{1}{4} \Tr\sigma_\nu \Tr\sigma_\mu \Tr \Lambda^4 = 0
	\end{equation}
	because the trace of Pauli matrices vanishes. For the variance one obtains
	\begin{equation}
		\mathbb{E}\left[ \left. \left(\partial_i S_2(\theta)\right)^2\right|_{\theta=0}\right] = \frac{8 M_A M_B}{(M_A^2-1)(M_B^2-1)} \left( e^{2(S_2 - S_3)} -1\right),
		\label{eq:variance}
	\end{equation}
	where $M_A=2\chi^{z_\nu-1}$, $M_B=2\chi^{z_\mu-1}$. Note that the above expression is always non-negative, as the R\'enyi entropy satisfies $S_2 \geq S_3$.  Setting $z_\nu=z_\mu=z$, $\chi \gg 1$, one gets
	\begin{equation}
		\mathbb{E}\left[ \left. \left(\partial_i S_2(\theta)\right)^2\right|_{\theta=0}\right] \approx  \frac{2}{\chi^{2(z-1)}} \left( e^{2(S_2 - S_3)} -1\right).
	\end{equation}
	That is, the variance decays only polynomially in the bond dimension $\chi$. The system size $N$ does not enter anywhere because we are dealing with optimization involving a local object of the form $Q(\theta)$.	Lastly, in the MPS case $M_A=M_B=2\chi$ the variance reduces to 
	\begin{equation}
		\mathbb{E}\left[ \left. \left(\partial_i S_2(\theta)\right)^2\right|_{\theta=0}\right] = \frac{32\chi^2}{(4\chi^2-1)^2} \left( e^{2(S_2 - S_3)} -1\right).
	\end{equation}
	
	\section{TFIM ground state search}
	\subsection{Imaginary time evolution in the BP-gauge}
	The TFIM ground states are obtained by imaginary-time evolution in a TEBD-like \cite{vidal_efficient_2003} scheme. For the Hamiltonian Eq.~(8) from the main text, written as ${H=\sum_{\mu,\nu}h_{\mu,\nu}}$ with $h_{\mu,\nu}$ being two-qubit terms on sites $\mu,\nu$, the non-unitary gates $e^{-\tau h_{\mu\nu}}$ are grouped into commuting layers by edge coloring; applying one gate updates the local messages, after which a full BP sweep (tolerance $10^{-6}$, at most $200$ iterations) is re-run. Each gated site tensor is renormalized by its Frobenius norm after every gate to prevent numerical overflow.
	
	In the ferromagnetic phase the finite-size ground state is a near-degenerate $\mathbb{Z}_2$ cat state. A small longitudinal field $h_z\sum_i Z_i$ is added during the early parts of the schedule, ramped from $h_z=0.05$ down to $10^{-4}$ and then switched off, which pins one ferromagnetic sector. Bond dimension is increased along the ladder $\chi=(2,4,8,16,32,64,128)$, each run initialized from the state already converged at the next-smaller $\chi$, to stabilize and speed up the convergence. The timestep $\tau$ drops from $\tau=0.1$ at the start to $\tau=0.001$ at the end of the evolution. We perform $1600$ timesteps in total. 
	
	\subsection{Comparison to different methods}
	\begin{figure}
		\includegraphics{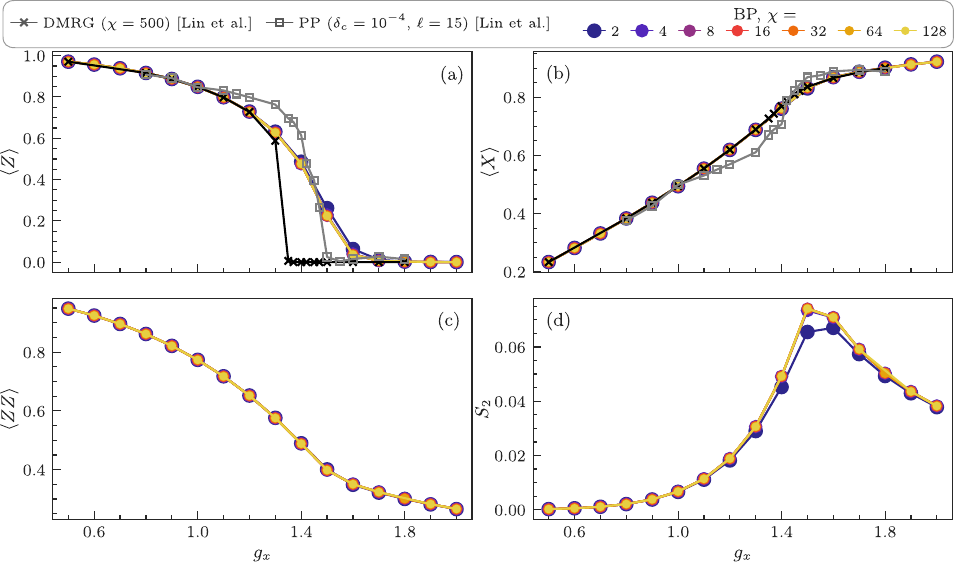}
		\caption{Properties of the obtained ground states as a function of the transverse field, compared to data from \cite{lin_utilityscale_2026}. All datapoints are averaged either over sites or over bonds.}
		\label{fig:smtfim1}
	\end{figure}
	\begin{figure}
		\includegraphics{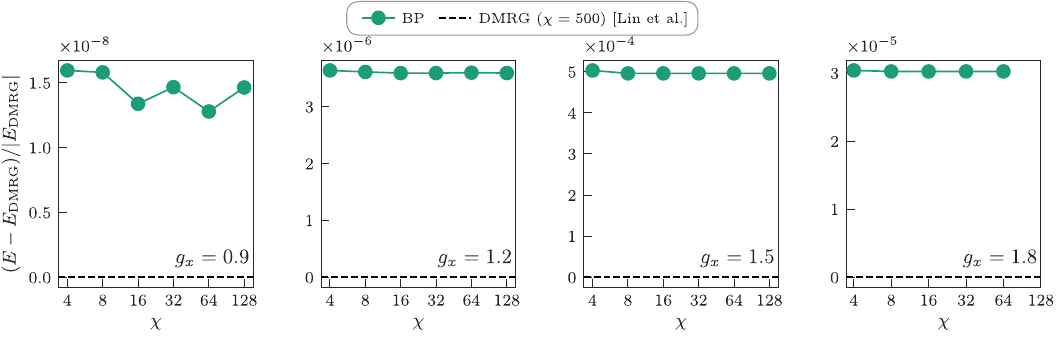}
		\caption{Differences of the ground state energies to the DMRG result of \cite{lin_utilityscale_2026} across the phase diagram as a function of the bond dimension.}
		\label{fig:smtfim2}
	\end{figure}
	In Fig.~\ref{fig:smtfim1} we plot the observables $\langle Z \rangle$, $\langle X \rangle$ (averaged over sites), $\langle ZZ \rangle$ and edge entropy $S_2$ (averaged over bonds) as a function of the transverse field $g_x$, obtained at different $\chi$. Clearly, the results are converged in the bond dimension $\chi$ (which, however, does not guarantee that we reach the exact solution over the entire phase diagram, as the calculations are done within the BP approximation, which disregards loop corrections). For state preparation demonstration in the main text, we selected only the ${\chi=32}$ case. 
	
	The observables $\langle Z \rangle$, $\langle X \rangle$ are compared to the data from \cite{lin_utilityscale_2026} which includes the DMRG calculation at ${\chi=500}$ and the Pauli-path propagation (PP) data at ${\delta_c=10^{-4}}$ truncation cutoff, ${\ell=15}$ circuit repetition. The order parameter $\langle Z \rangle$, Fig.~\ref{fig:smtfim1}(a), reveals a ferro- to paramagnet transition in all cases but at varying critical $g_x$. Our data closely reproduces the DMRG result apart from the region close to the transition, which can be related to the BP's inability to capture long-range correlations at critical points \cite{midha_belief_2026}. On the other hand, the $\langle X \rangle$ observable from BP, Fig.~\ref{fig:smtfim1}(b), matches with DMRG very well across the entire phase diagram. The PP result, as the authors suggest in private communication, should be interpreted with caution given the finite, possibly too large, truncation threshold used. The $\langle ZZ \rangle$, Fig.~\ref{fig:smtfim1}(c), and $S_2$, Fig.~\ref{fig:smtfim1}(d), do not change when increasing the bond dimension above $\chi\geq 4$. In Fig.~\ref{fig:smtfim2} we compare the obtained energies to the DMRG energies, finding a good relative agreement and no strong drift with the bond dimension $\chi$.

	\input{supplemental.bbl}

%% file: main.bbl
%

%% file: supplemental.bbl
%

%% file: main.bbl
\begin{thebibliography}{107}
\bibitem [{foo({\natexlab{b}})}]{footnote4}%
  \BibitemOpen
  \href@noop {} {\bibinfo {title} {Messages are initialized as positive
  semi-definite and stay as such during iterations of {E}q. (1), so the square
  root always exists. {In} our calculations, the messages are initialized as
  identity matrices.}} \BibitemShut {NoStop}%
\bibitem [{foo({\natexlab{d}})}]{footnote3}%
  \BibitemOpen
  \href@noop {} {\bibinfo {title} {A bipartite graph is a graph which does not
  contain any odd-length cycles.}} \BibitemShut {NoStop}%
\end{thebibliography}
